\documentclass[superscriptaddress,twocolumn,aps,pra,amsmath,floatfix]{revtex4}
\usepackage{graphicx}
\begin{document}

\title{Surface vs. bulk Coulomb correlations in photoemission spectra 
           of perovskites}

\author{A. Liebsch}
\affiliation{Institut f\"ur Festk\"orperforschung, Forschungszentrum
             J\"ulich, 52425 J\"ulich, Germany}
\date{\today }

\begin{abstract}
Recent photoemission spectra of the perovskite series
Sr$_x$Ca$_{1-x}$VO$_3$ revealed strong modifications associated with
surface contributions. To study the effect of Coulomb correlations
in the bulk and at the surface the quasi-particle spectra are evaluated  
using the dynamical mean field theory. It is shown that as a result of 
the reduced coordination number of surface atoms correlation effects 
are stronger at the surface than in the bulk, in agreement with 
experiment.\\   

\end{abstract}

\maketitle

Photoemission is a key spectroscopy for the investigation
of electronic properties of strongly correlated materials. Because
of its surface sensitivity there is growing concern to what 
extent the data are representative of bulk properties and 
whether they need to be corrected for surface effects influencing the
one- and many-electron properties. To study these questions Maiti 
{\it et al.}\ \cite{maiti} and Sekiyama {\it et al.}\ 
\cite{sekiyama} performed    
photoemission measurements on Sr$_x$Ca$_{1-x}$VO$_3$ using a 
wide range of photon energies. As a result of the frequency dependent 
escape depth of the photoelectron, the spectra reveal striking 
variations evidently associated with surface modifications of the 
electronic structure: emission from the coherent peak near the 
Fermi level is reduced and correlation-induced satellites have larger 
weight than in the bulk. The same trend is observed on
SrRuO$_3$\ \cite{fujioka}. Another example is Sr$_2$RuO$_4$ for which 
earlier photoemission data appeared to be in conflict with de Haas-%
van Alphen measurements\ \cite{SrRuO1}. Recent experimental and theoretical 
work proved that this discrepancy can be resolved by accounting for 
the surface reconstruction of Sr$_2$RuO$_4$ which leads 
to significant changes in photoemission spectra\ \cite{SrRuO2}. 

The understanding of surface effects in photoemission spectra of 
transition metal oxides is important in order to distinguish them 
from correlation phenomena in the bulk. Although the electronic 
properties of strongly correlated materials are currently a field of 
intense experimental and theoretical investigation\ 
\cite{imada,georges,held} 
surface effects have so far received little attention\ \cite{potthoff}. 
For this purpose photoemission studies of the series
Sr$_x$Ca$_{1-x}$VO$_3$\ \cite{maiti,sekiyama,fujimori,morikawa,inoue} 
are very useful since the one-electron properties are relatively simple 
and the ratio $U/W$ can be tuned systematically. Here, $U$ is the 
on-site Coulomb energy and $W$ the width of the relevant bands near the 
Fermi level. While SrVO$_3$ is considered to be a moderately correlated 
metal with a clear quasi-particle peak near $E_F$, the slightly smaller
band width of CaVO$_3$ is believed to push this system closer 
to the metal--insulator transition. For this reason CaVO$_3$ has become a 
key system of investigation\ \cite{fukushima,shirakawa,inoue02}.        
For instance, recent de Haas-van Alphen measurements
on CaVO$_3$\ \cite{inoue02} detected a cubic Fermi surface although 
this system exhibits orthorombic distortions. 

In the present work we use the dynamical mean field theory (DMFT)
\ \cite{georges,vollhardt,pruschke} to evaluate the surface and bulk
quasi-particle spectra of SrVO$_3$. We show that because of the 
planar electronic structure of perovskite systems and the concomitant 
narrowing of the surface local density of states correlation 
effects at the surface are more pronounced than in the bulk.
In the perovskites this enhancement is particularly strong since 
the local Coulomb energy is not far from the critical value for a 
metal-insulator transition. Such a
trend had been predicted by Potthoff and Nolting\ \cite{potthoff}
who studied the metal--insulator phase diagram for a semi-infinite 
simple cubic $s$ band at half filling. Here we calculate the self-energy  
for a multi-band system using a realistic local density of states 
appropriate for SrVO$_3$ and find qualitative agreement with 
photoemission data\ \cite{maiti,sekiyama}. 

Electronic structure calculations for SrVO$_3$ within the local density 
approximation (LDA)\  \cite{takegahara} show that the conduction bands 
near $E_F$ consist of degenerate $t_{2g}$ bands derived from V$^{4+}$ 
($3d^1$) ions. The filled O 2p bands are separated from the $t_{2g}$ 
levels by a gap of about 1~eV, and the cubic crystal field of the V-O 
octahedron shifts the $e_g$ bands above the $t_{2g}$ bands. The $t_{2g}$ 
bands can be represented via a tight-binding Hamiltonian with diagonal 
elements
\,$h_{xy,xy}(k)= e_d + t_0 (c_x + c_y) + t_1 c_x c_y 
              + [t_2  + t_3(c_x + c_y) + t_4 c_x c_y] c_z$,       
where $c_i = 2\cos(k_ia),\, i=x,y,z$ and $a$ is the lattice constant. 
Cyclic permutations yield $h_{xz,xz}(k)$ and $h_{yz,yz}(k)$.
The $t_i$ denote effective hopping integrals representing the V-O-V 
hybridization, where $t_{0,2}$, $t_{1,3}$, and $t_4$ specify the 
interaction between first, second and third neighbors, respectively.
For symmetry reasons off-diagonal elements arise only between second 
and third nearest neighbors and are of the form \,$h_{xy,xz}(k) = 
- t'_1 s_y s_z - t'_2 c_x s_y s_z$, where $s_i = 2\sin(k_ia)$, i.e.,   
they vanish at high-symmetry points. Since the $t'_{1,2}$ are very 
small we neglect the off-diagonal elements so that the energy bands are 
given by \,$\epsilon_i(k) =  h_{i,i}(k)$, with \,$i=xy,xz,yz$.
The parameters $e_d$ and $t_0\cdots t_4$ can be found by fitting the 
LDA energies at high-symmetry points of the Brillouin Zone. 
 
\begin{figure}
  \begin{center}
  \includegraphics[width=5cm,height=8cm,angle=-90]{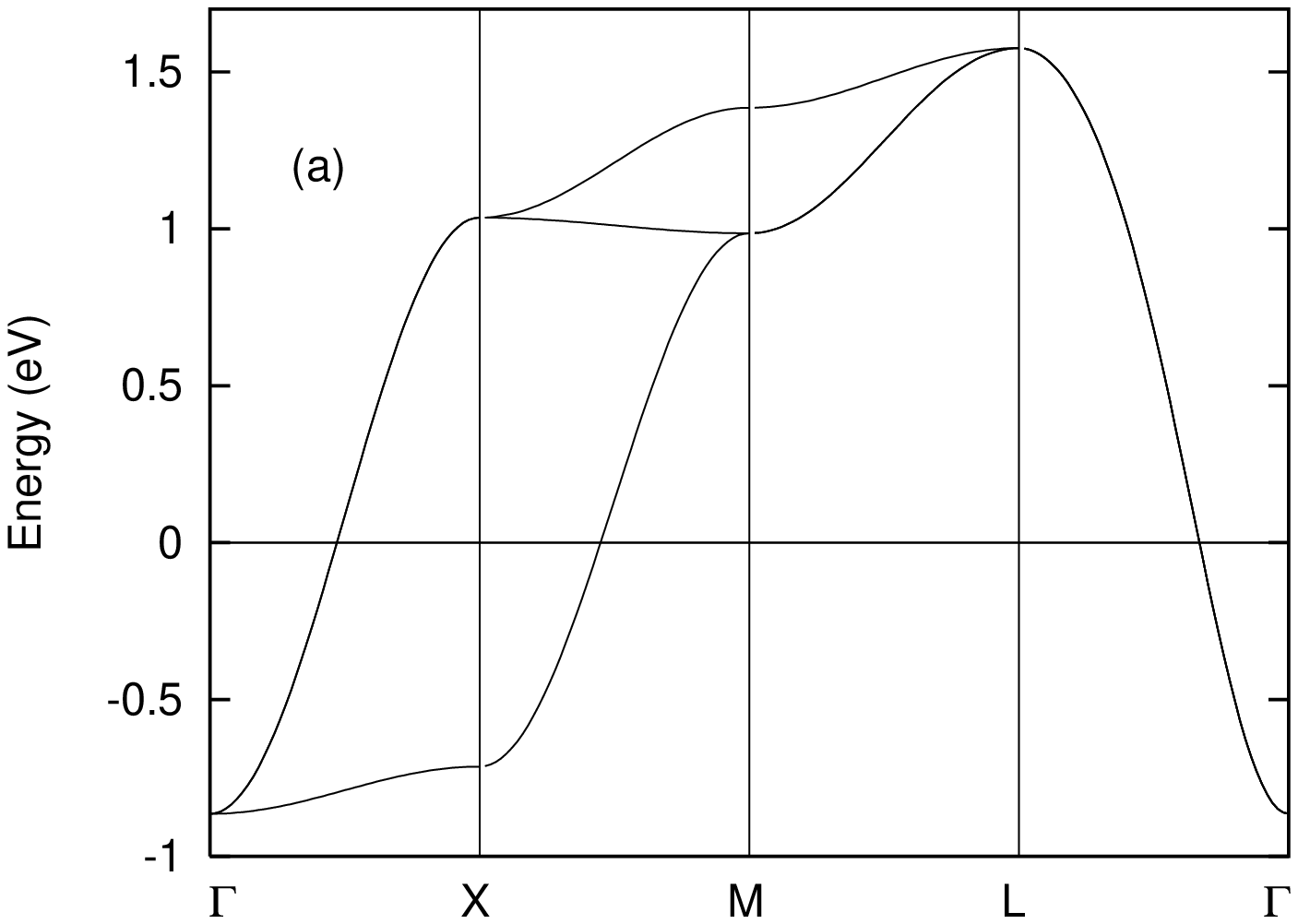}
  \includegraphics[width=5cm,height=8cm,angle=-90]{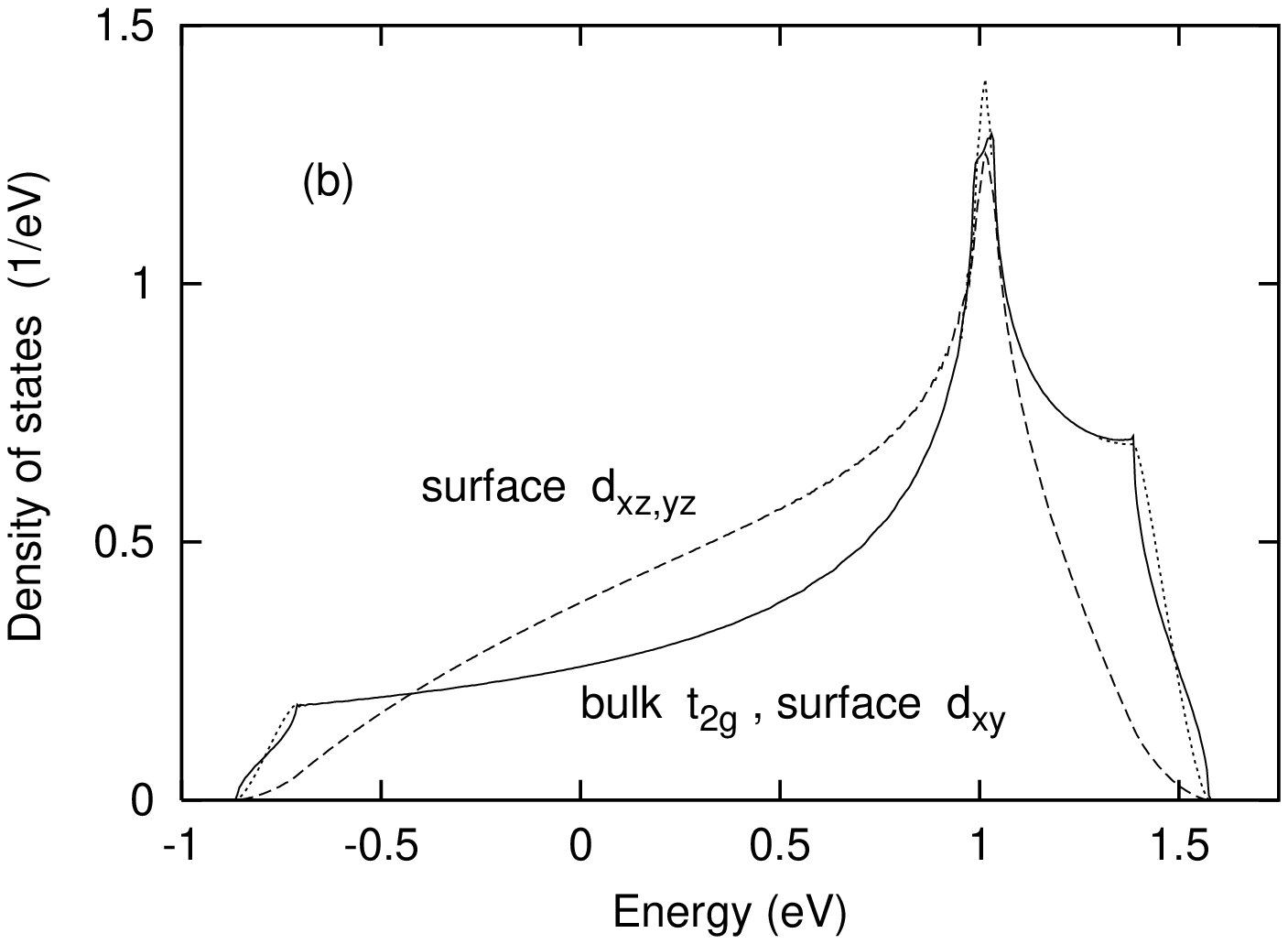}
   \end{center}
\vskip-0.4cm
\caption{
(a) Tight-binding fit to LDA  $t_{2g}$ bulk bands of SrVO$_3$ ($E_F=0$). 
(b) Solid curve: isotropic bulk density of states;
dashed and dotted curves: local density of $d_{xz,yz}$ and $d_{xy}$ 
states in the first layer of SrVO$_3$. 
}\end{figure}

Fig.~1\,(a) shows the $t_{2g}$ bulk bands of SrVO$_3$. According to the
predominantly planar electronic structure, the bulk density of states 
$\rho_b(\omega)$ exhibits the characteristic asymmetric peak related 
to the van Hove singularity at the X point (see Fig.~1\,(b)). The overall 
shape of the bulk density of states agrees with LDA calculations\ 
\cite{takegahara}. At the surface, the $t_{2g}$ degeneracy is lifted 
since only the $d_{xy}$ band exhibits strong dispersion within 
the surface plane (the $z$ axis defines the surface normal).
To evaluate the surface density of states we use a Green's function 
formalism\ \cite{kalkstein} for semi-infinite tight-binding systems.
As electronic structure calculations for SrVO$_3$ surfaces are not 
available, we assume the tight-binding parameters to coincide with 
those in the bulk, except for a small surface potential to ensure 
charge neutrality.

Fig.~1\,(b) shows the local density of states $\rho_s(\omega)$ 
of the $d_{xz,yz}$ bands for the surface layer of  SrVO$_3$. Its 
spectral weight is reduced at low and high energies but enhanced at 
intermediate energies. Thus, the effective width of $\rho_s(\omega)$ 
is smaller than that of $\rho_b(\omega)$ although their total widths 
are identical. In contrast to the $d_{xz,yz}$ density, the density 
of $d_{xy}$ states nearly coincides with the bulk density, reflecting
the planar nature of the $t_{2g}$ states. In the deeper layers
the local density of $d_{xz,yz}$ states approaches $\rho_b(\omega)$
rather quickly, the main effect consisting in an oscillatory distortion 
of the spectral shape. This rapid convergence is a consequence of the 
tight-binding character of the $t_{2g}$ bands. It would be desirable 
to carry out self-consistent electronic structure calculations for 
SrVO$_3$ surfaces since they should provide a more accurate density 
of states. Nevertheless, the key effect discussed here, namely, the 
preferential narrowing of the local density of $d_{xz,yz}$ states, 
should hold quite generally.             

To interpret the experimental photoemission data we evaluate the 
quasi-particle spectra by accounting for local Coulomb correlations.
According to the one-electron properties discussed above we are 
dealing with a non-isotropic system where two narrow bands interact 
with a wider band. In the first layer this difference in 
effective band width is largest and it diminishes rapidly towards 
the interior of the system. This situation is reminiscent of the layer 
perovskite Sr$_2$RuO$_4$, which essentially consists of Ru sheets 
containing two narrow $d_{xz,yz}$ bands interacting with a wide 
intra-planar $d_{xy}$ band. The peculiar feature of this system is 
the fact the on-site Coulomb energy lies between the single-particle 
widths of the non-degenerate $t_{2g}$ bands: \,$W_{xz,yz} < U < W_{xy}$
\ \cite{liebsch}. In the present case the difference 
between $d_{xz,yz}$ and $d_{xy}$ bands at the surface is less 
pronounced so that $W_i<U$ for all three bands. Nevertheless, since 
in the first layer the effective width of $d_{xz,yz}$ states is 
reduced, the influence of Coulomb correlations on the surface bands 
should be stronger than on the wider bulk bands.   
 
The key quantity characterizing the effect of Coulomb correlations 
is the complex self-energy. Since we neglect the 
weak hybridization between $t_{2g}$ bands, $\Sigma(\omega)$ is 
diagonal in orbital space. Despite this simplification, a full 
description of correlations near the surface would be exceedingly 
complicated since, in principle, it would require a mixed momentum/real 
space approach in order to handle the loss of translational symmetry 
normal to the surface. This could be accomplished using a cluster 
formalism where the semi-infinite system is represented via a 
slab of finite thickness. Unfortunately, the planar character of
$t_{2g}$ orbitals leads to a slowly convergent local density of 
$d_{xz,yz}$ states, with many spurious $1/\sqrt{\omega}$ van 
Hove singularities stemming from the quasi-one-dimensional hopping
parallel to the surface. On the other hand, a cluster generalization 
of the DMFT is feasible today only for very small cluster size.
We therefore ignore the momentum variation of the self-energy and 
assume that, for a given layer, it depends only on the local density
of states within that layer\ \cite{potthoff}. The reduced
dimensionality at the surface enters the many-body calculation  
via the layer-dependent density of states. In view of the local 
Coulomb interaction this assumption seems justified for qualitative 
purposes.

Before discussing the quasi-particle spectra derived within the DMFT
it is useful to compare the frequency dependence of the self-energy
in the bulk and at the surface. For convenience we give here the 
results obtained within non-self-consistent second-order perturbation 
theory since this provides a clear identification of the various orbital 
contributions to the diagonal elements $\Sigma_i(\omega)$\ \cite{liebsch}.
Fig.~2 illustrates the second-order self-energy for 
$U=4.3$~eV and $J=0.7$~eV. Although the total widths of the bulk 
and surface densities shown in Fig.~1 are the same, the 
effective band narrowing at the surface yields a correspondingly 
narrower distribution of Im\,$\Sigma_{xz,yz}(\omega)$ compared to 
Im\,$\Sigma_b(\omega)$ for the isotropic bulk.
Because of the orbital mixing via the on-site Coulomb interaction 
Im\,$\Sigma_{xy}(\omega)$ lies between these two distributions.
From the Kramers-Kronig relations it then follows that  
Re\,$\Sigma_{xz,yz}(\omega)$ exhibits a more pronounced 
minimum below the band bottom than Re\,$\Sigma_b(\omega)$. 
Thus, as indicated by the line $\omega-\epsilon_k$ for a typical
band energy $\epsilon_k$, the surface satellite appears at a 
lower value of $U$ than in the bulk. Analogous arguments hold for
the upper Hubbard peak. This qualitative picture is fully supported 
by the more accurate DMFT results discussed below. 
 
\begin{figure}
  \begin{center}
  \includegraphics[width=5.0cm,height=8cm,angle=-90]{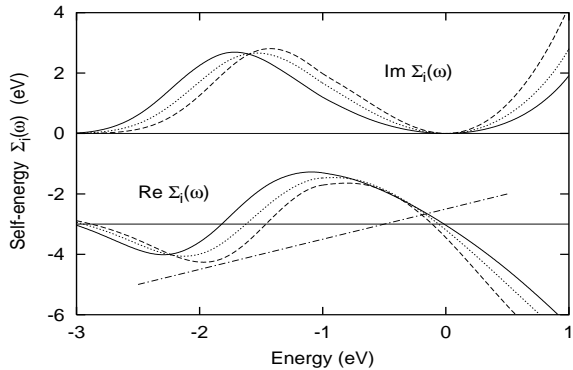}
   \end{center}
\vskip-0.4cm
\caption{
Second-order self-energy of SrVO$_3$.
Solid curves: bulk $t_{2g}$ self-energy; 
dashed (dotted) curves: $d_{xz,yz}$ ($d_{xy}$) self-energy
for surface layer. For clarity the real part is shifted down 
by 3~eV. The straight line gives $\omega-\epsilon_k$ for 
$\epsilon_k=-0.5$~eV.
}\end{figure}

In the DMFT the self-energy elements $\Sigma_i(\omega)$  
are functionals of the bath Green's functions 
\,${\cal G}_i^{-1} = G_i^{-1} + \Sigma_i$.
The quasi-particle densities of states are related to the local $G_i$
via \,$N_i(\omega)=-{\rm Im}\,G_i(\omega)/\pi$.
On-site Coulomb correlations are treated using the self-consistent 
multi-band Quantum Monte Carlo method (for a review, see Ref.\ 
\cite{georges}). The temperature of the simulation is 12.5~meV with 
64 imaginary time slices and several runs using $10^5$ Monte Carlo 
sweeps. The quasi-particle density of states is obtained via maximum 
entropy reconstruction\ \cite{jarrell}.
 
\begin{figure}
\begin{center}     
  \includegraphics[height=8cm,width=5.0cm,angle=-90]{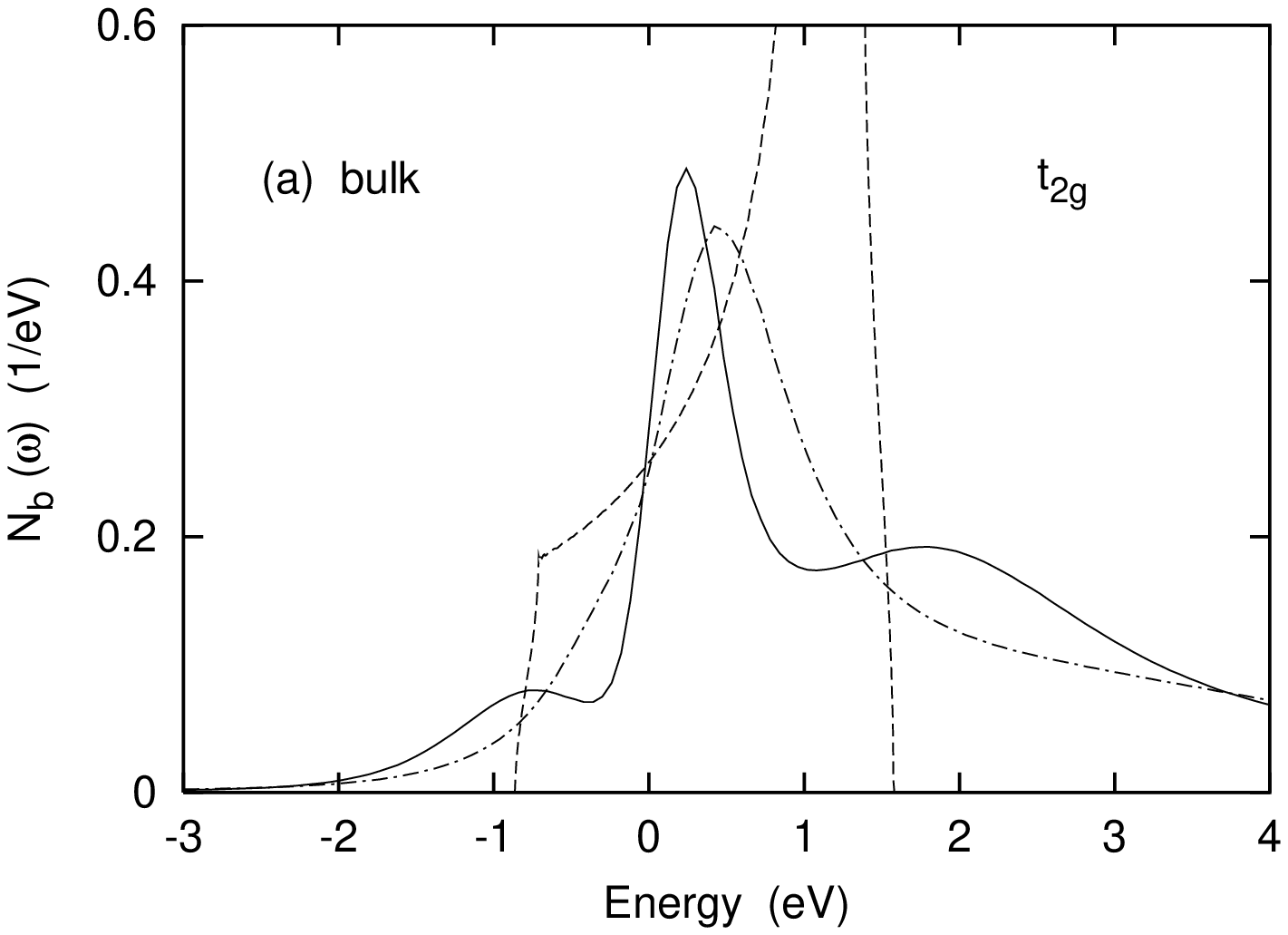}
  \includegraphics[height=8cm,width=5.0cm,angle=-90]{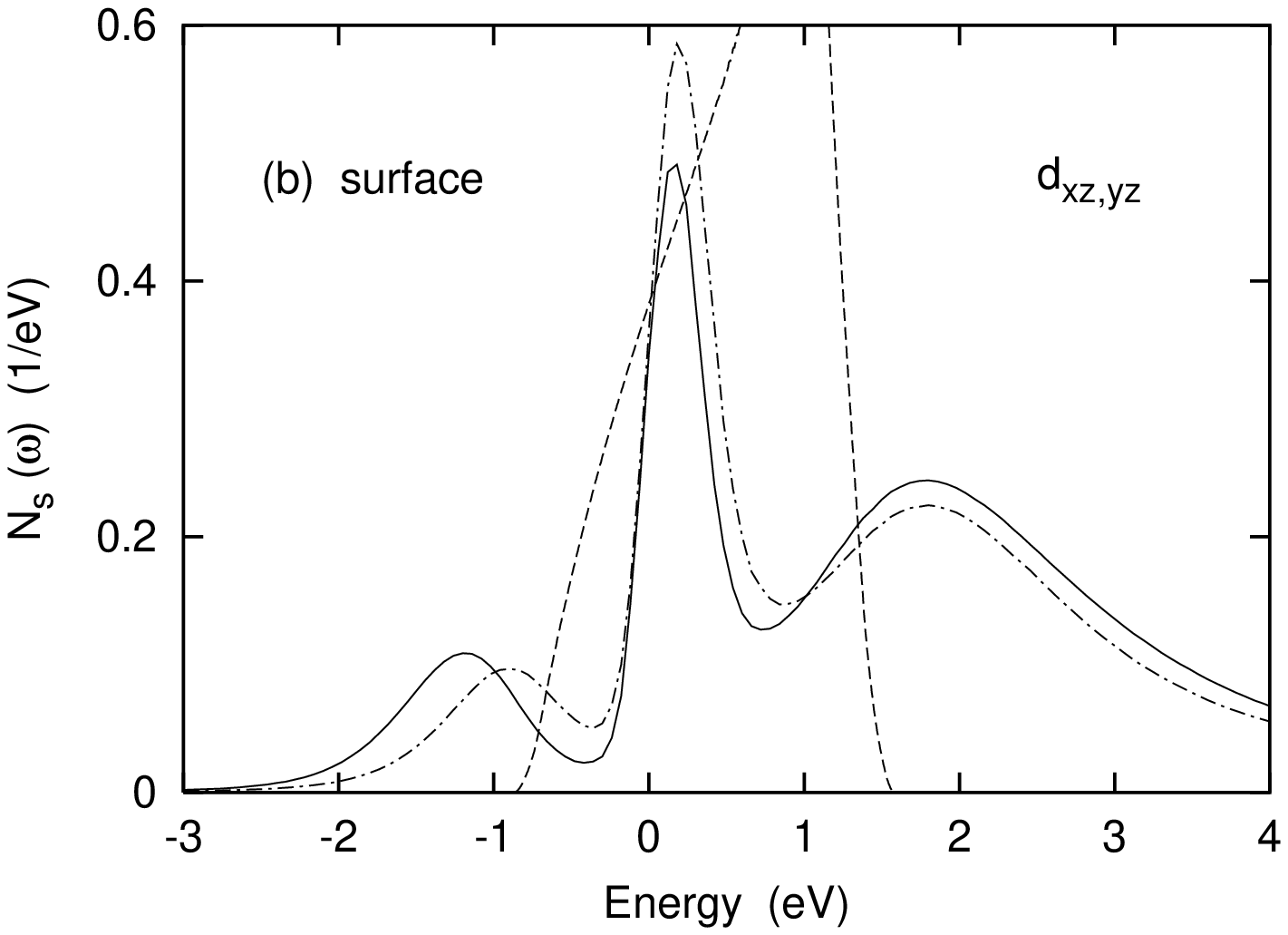}
  \includegraphics[height=8cm,width=5.0cm,angle=-90]{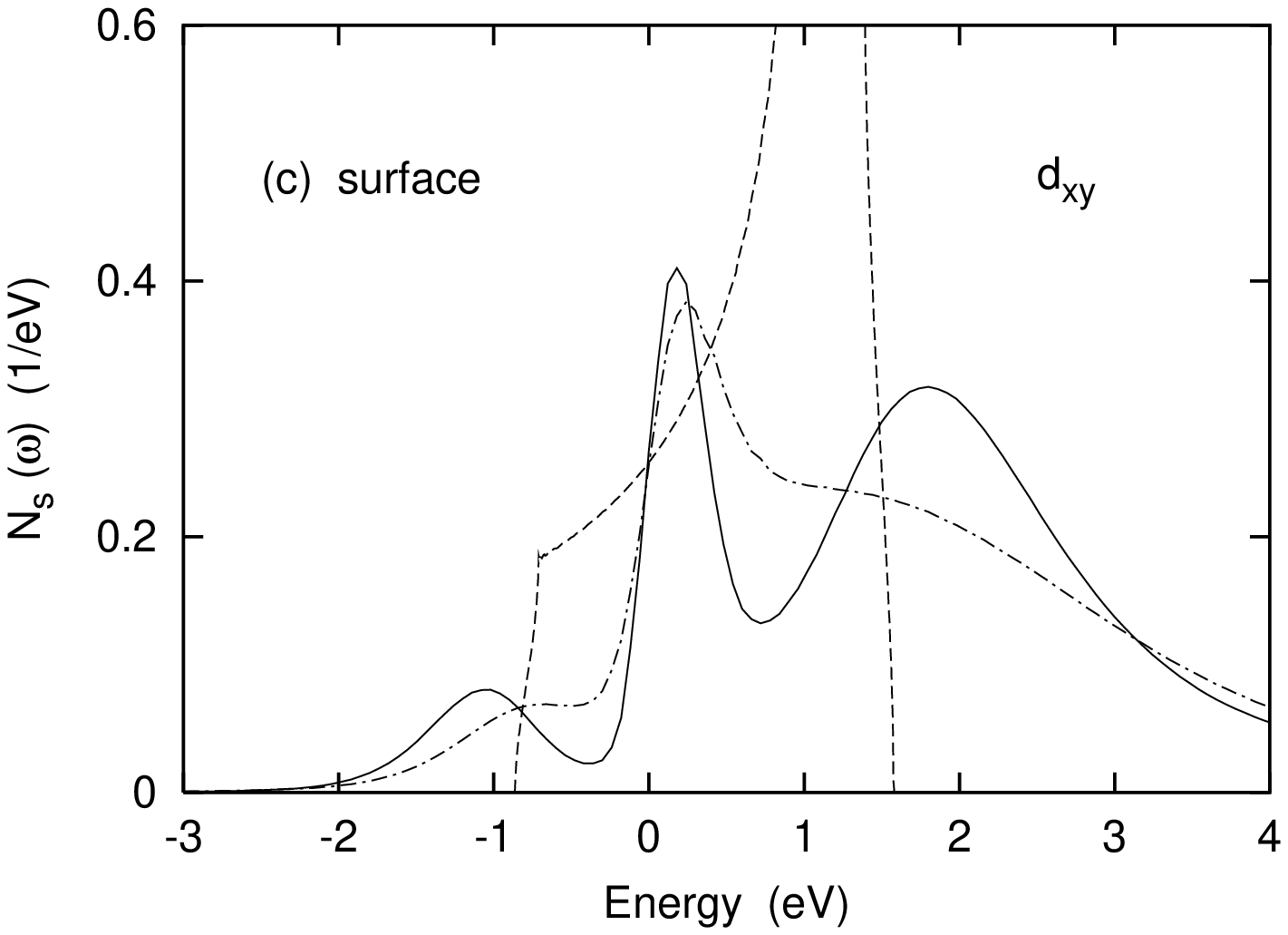}
\end{center}
\vskip-0.4cm
\caption{
Quasi-particle density of states of SrVO$_3$ derived from DMFT.
(a) bulk $t_{2g}$ states,
(b) surface $d_{xz,yz}$ states, 
(c) surface $d_{xy}$    states. 
Solid curves: $U=4.3$\,eV; dot-dashed curves: $U=4.0$\,eV; 
dashed curves: bare densities.
}\end{figure}

Fig.~3\,(a) shows the bulk quasi-particle density of states of  
SrVO$_3$  for two Coulomb energies:  $U=4.0$\,eV and  
$4.3$\,eV. The exchange energy is $J=0.7$\,eV\ \cite{zaanen}. 
These results show that in the bulk $U$ must be larger than 4\,eV
to obtain the satellite observed in photoemission spectra 
\cite{maiti,sekiyama}. The peak near 2~eV above $E_F$ agrees with 
inverse photoemission data\ \cite{morikawa}. Although for $U=4$\,eV 
there is considerable correlation-induced band narrowing and an 
emerging satellite shoulder, the increase of $U$ to $4.3$\,eV yields 
an even narrower coherent feature near $E_F$, with the missing weight 
shifted to the lower and upper Hubbard bands. Note that 
$N_b(E_F)=\rho_b(E_F)$ which follows from the local approximation
implicit in the DMFT for isotropic systems\ \cite{pinning}. The most 
recent photoemission data\ \cite{sekiyama} confirm this result. The 
spectra shown in Fig.~3\,(a) qualitatively agree with DMFT-LDA results 
for the $3d^1$ perovskite La$_x$Sr$_{1-x}$TiO$_3$\ \cite{nekrasov}
which exhibits a similar $t_{2g}$ bulk density of states.  

The surface quasi-particle spectra for SrVO$_3$ are shown in 
Fig.~3\,(b) and (c). The lower Hubbard peak of the $d_{xz,yz}$ states 
is clearly visible already for $U=4$\,eV because of the narrower local 
density of states in the first layer. A larger $U$ shifts the 
satellite to higher binding energies. The comparison with the spectra
shown in Fig.~3\,(a) demonstrates that correlation effects for a fixed 
value of $U$ are stronger at the surface than in the bulk: The coherent 
peak near $E_F$ is narrower at the surface and the incoherent satellite 
feature is more pronounced than in the bulk, in agreement with 
experiment\ \cite{maiti,sekiyama,fujioka}. 

The surface quasi-particle density of $d_{xy}$ states in Fig.~3\,(c) 
is intermediate between $N_b(\omega)$ and $N_s(\omega)$ for 
$d_{xz,yz}$. Although there is little single-electron hybridization 
between $t_{2g}$ bands, the local Coulomb interaction mixes them so that 
the $d_{xy}$ surface spectrum involves contributions arising from the 
more strongly correlated $d_{xz,yz}$ states. Another consequence
of the anisotropic surface self-energy is the 
correlation-induced charge transfer between subbands\ \cite{liebsch}. 
Here, we find that 0.06 electrons are shifted from the $d_{xy}$ to 
the $d_{xz,yz}$ bands. Also, the quasi-particle partial densities of 
states at $E_F$ do not need to coincide with the bare partial densities. 
The coupling between narrow and wide bands is a genuine multi-band effect 
and underlines the fact that single-particle bands in the presence of 
local Coulomb interactions cannot be considered independently. This 
issue will be considered in more detail elsewhere\ \cite{future}.  

Note that the many-body reduction of the quasi-particle band width is 
much larger than the surface-induced one-electron band narrowing. 
On the other hand, since with $U/W\approx2$ the on-site Coulomb 
energy is not far from the critical value for a metal-insulator 
transition \cite{georges}, the band narrowing substantially 
enhances the influence of correlations at the surface.   

It would be interesting to perform angle-resolved photoemission 
measurements to determine the correlation-induced band narrowing  
of the SrVO$_3$ $t_{2g}$ bands. For instance the energy at $\bar\Gamma$ 
should be only a few tenths of an eV below $E_F$ instead of 1\,eV 
as predicted by the LDA. Accordingly, the true $t_{2g}$ bands should be 
considerably flatter than the single-particle bands. Also, measurements 
using polarized light could help to separate correlation effects in the 
$d_{xy}$ and $d_{xz,yz}$ bands.
 
We finally comment on the difference between CaVO$_3$ and SrVO$_3$.
According to LDA band structure calculations for CaRuO$_3$ and 
SrRuO$_3$\ \cite{mazin}, orthorombic distortions cause a slight 
narrowing of the $t_{2g}$ bands and a strong broadening of the van 
Hove singularity above $E_F$. Assuming the same on-site Coulomb energy 
for the V$^{4+}$ ions in CaVO$_3$ and SrVO$_3$, a smaller bulk band 
width of CaVO$_3$ implies a transfer of spectral weight from the 
coherent to the incoherent peak. At present, the available data\ 
\cite{maiti,sekiyama,fujimori,morikawa,inoue}
do not yet fully agree on the magnitude of this effect.    
At the surface, such a transfer of spectral weight should be more 
pronounced because of the band narrowing discussed in the present paper.
This effect is indeed observed in the photoemission spectra\ 
\cite{maiti,sekiyama}. For a more detailed analysis it would be 
important to carry out electronic structure calculations for  
CaVO$_3$ in order to study orthorombic distortions
and possible reconstructions  at the surface. 

In summary, we have performed DMFT quasi-particle calculations for 
SrVO$_3$ in the bulk and at the surface. As a result of the planar 
nature of the $t_{2g}$ states and the surface narrowing of the local 
density of states on-site correlation effects are more pronounced at 
the surface than in the bulk, in agreement with photoemission data. 
The two-dimensional character of $d$ states near 
$E_F$ is one of the hallmarks of transition metal oxides. The 
surface-induced enhancement of correlation effects discussed in the 
present work should therefore be a phenomenon observable in many 
materials.   

I thank A. Bringer for discussions and A.I. Lichtenstein 
for the QMC-DMFT code. I also thank S. Biermann for comments and for 
bringing Ref.~2 to my attention.

\end{document}